\documentclass{ifacconf}

\usepackage[dvipdfmx]{graphicx}     % include this line if your document contains figures
\usepackage{natbib}        % required for bibliography
\usepackage{mathtools}
\usepackage{amsmath, amssymb}
\mathtoolsset{showonlyrefs,showmanualtags}
\usepackage{url}
\usepackage{color}

\newcommand{\rfig}[1]{Fig.~\ref{#1}} 
\newcommand{\Ucal}{\mathcal{U}}
\newcommand{\Ecal}{\mathcal{E}}

\newcommand{\Rn}{\mathbb{R}^n}

\newtheorem{assu}{Assumption}
\newtheorem{theo}{Theorem}
\newtheorem{rema}{Remark}

\newtheorem{problem}{Problem}

\begin{document}
\begin{frontmatter}

\title{Weak Control Approach to Consumer-Preferred Energy Management}
%Approach to Energy Management System with Allowing of Consumers' Selfish Decision 

\thanks[footnoteinfo]{This work was supported by CREST through JST under Grant JPMJCR15K1.  S. Shibasaki and M. Inoue contributed equally to this work.}

\author[First]{Suzuna Shibasaki}
\author[First]{Masaki Inoue}
\author[First]{Mitsuru Arahata} 
\author[Second]{Vijay Gupta} 
%vijay

\address[First]{Department of Applied Physics and Physico-Informatics,\\
Keio University, 
3-14-1 Hiyoshi, Kohoku-ku, Yokohama, Kanagawa, Japan
(e-mail: suzuna.s@keio.jp, minoue@appi.keio.ac.jp, arahatamitsuru@keio.jp)}
\address[Second]{Department of Electrical Engineering, University of Notre Dame, 
275 Fitzpatrick Hl Engrng, Notre Dame, IN, USA
(e-mail: vgupta2@nd.edu)}

\begin{abstract}                % Abstract of not more than 250 words.
This paper is devoted to a consumer-preferred community-level energy management system (CEMS), 
in which a system manager allows consumers their selfish decisions of power-saving while regulating the overall demand-supply imbalance.
The key structure of the system is to {\it weakly} control consumers: the controller sends the allowable range of the power-saving amount to each consumer, which is modeled by
a set-valued control signal.  Then, the consumers decide the amount in the range based on their private preference.  
In this paper, we address the design problem of the controller that generates the set-valued control signals. 
The controller structure is based on internal model control, which plays the essential role of guaranteeing 
the consumer-independent stability and the worst-case control performance of the overall CEMS.
Finally, a numerical experiment of the consumer-preferred CEMS is performed to demonstrate the design procedure of the controller and to show its effectiveness.
\end{abstract}

\begin{keyword}
Human-in-the-loop system, Energy management system, Multi-objective control, Internal model control, Robust control.
%Five to ten keywords, preferably chosen from the IFAC keyword list.
\end{keyword}

\end{frontmatter}
%=============================================================================
\section{Introduction}%Section1
The renewable energy has gained much importance in recent years due to its benefits of reducing power generation costs and carbon emission levels.  
To fully receive such benefits, the drawbacks of the renewable energy must be overcome. %by control systems technologies.  
For example, solar power or wind power generation fluctuates uncertainly depending on the weather condition.
The fluctuation may cause the power demand-supply imbalance and lead the blackout in the worst case. % (see e.g., the survey by \cite{haes2019survey}).  
Energy management technologies of balancing the demand and supply must be further developed.

One of the highly potential technologies is demand side management (DSM)
 (see e.g., the works by \cite{strbac2008demand},  \cite{palensky2011demand},  \cite{logenthiran2012demand}, \cite{khalid2018towards}).
DSM includes not only the direct management of mechatronic devices, but also indirect one using demand response (DR) of consumers.
 In DR, dynamic pricing, incentives, and other control methods promote consumers to change their power-usage based on their preference.
Such DR methods are under intense investigation in various research fields 
in the works  by e.g., \cite{albadi2008summary}, \cite{mohsenian2010optimal}, \cite{caron2010incentive}, \cite{siano2014demand}, \cite{qureshi2014model}, \cite{rahmani2016modeling}, \cite{dobakhshari2018contract}, \cite{he2018practical}, \cite{miyazaki2019design}.

It is desirable for DSM that consumers always respond to power-saving requests, 
which are given in the form of prices or incentives,  to achieve any target power-saving amount. 
However, consumers do not, and they may make selfish decisions of the power-saving amount based on their own private preference.
Any DR method unavoidably causes the error in the target power-saving amount and the actual amount.  A promising way to effectively reduce the error is to take the feedback structure into DR as studied by \cite{qureshi2014model}, \cite{he2018practical}, \cite{miyazaki2019design}.  
Feedback DR particularly taking care of irrational or selfish behavior of consumers, is addressed in this paper.

\begin{figure}[t]
\begin{center}
\includegraphics[width=\linewidth]{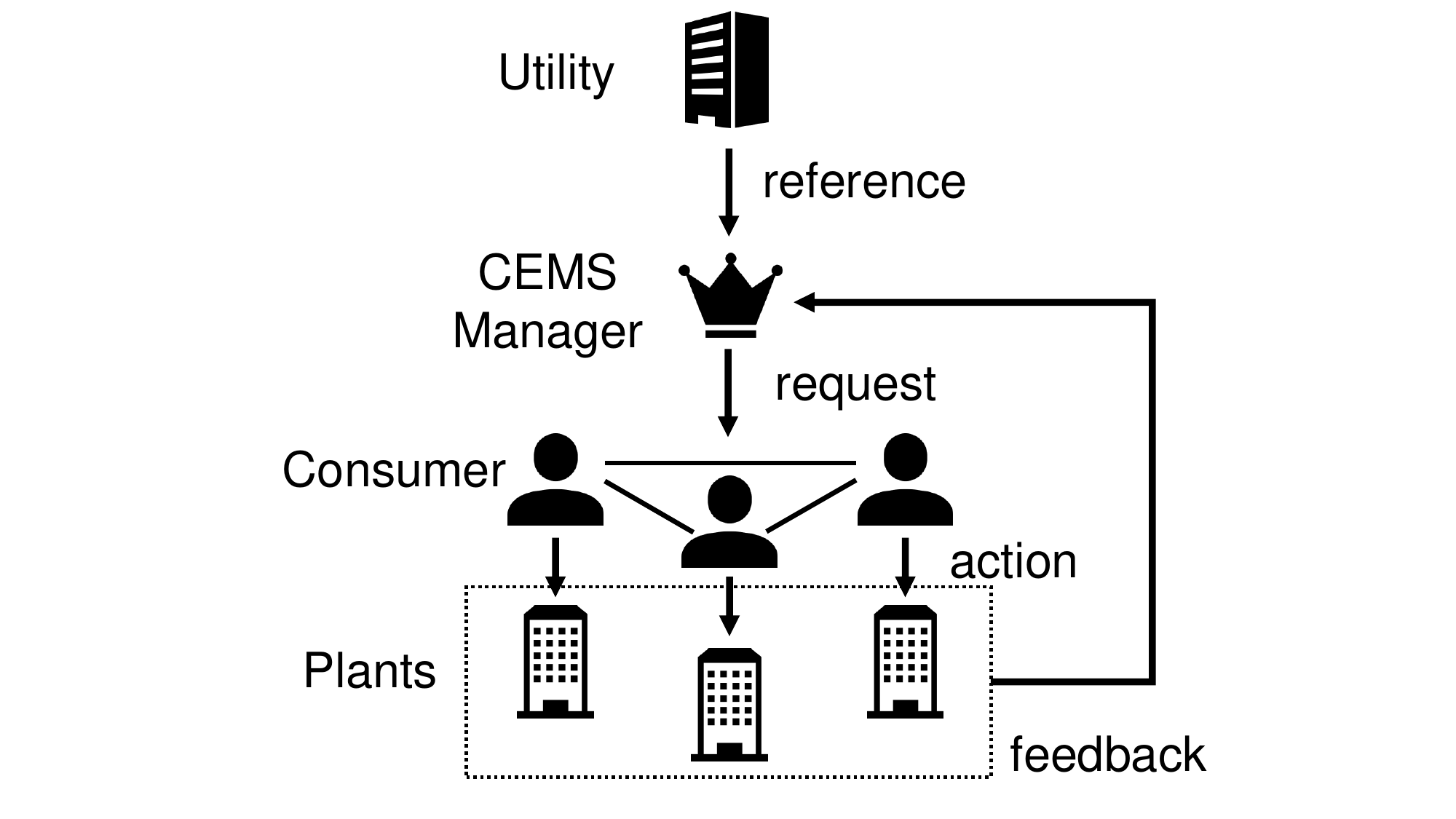}    % The printed column width is 8.4 cm.
\caption{Consumer-preferred CEMS} 
\label{block_CEMS}
\end{center}
\end{figure}

This paper addresses the feedback structure in a community-level energy management system (CEMS), which is illustrated in \rfig{block_CEMS}.
In the figure, a power utility sends the reference of power-saving amount to the CEMS manager, 
and the manager provides some request to consumers.
Then, the consumers take their power-saving actions to their plant systems individually.
In the CEMS, we particularly pursue the following two aims:  
one is the reliable feedback control of accurately achieving any target power-saving amount.
The other is the consumer-preferred structure of allowing consumers their selfish decisions of power-saving amount.

The key to realize the consumer-preferred structure is {\it weak} control, the concept of which is originally proposed by \cite{inoue2019weak}:
the request sent from the manager to the consumers is given by the {\it allowable range} of power-saving amount, which is modeled by a {\it set-valued} signal.
Then, the consumers decide the power-saving amount based on their own private preference.  The decision can be made without taking care of the stability or control performance of the overall CEMS.
In this paper, we address the design problem of the controllers that generate the set-valued signals.
The controller structure is based on internal model control (IMC, see e.g., the book by \cite{morari1989robust}).
The consumer-independent stability and the worst-case control performance of the overall CEMS are studied.
Finally, a numerical experiment of the consumer-preferred CEMS is performed to demonstrate the design procedure of the controller and to show its effectiveness.

%

%In the CEMS we consider in this paper, allowable action set is given to consumers.
%Then, each consumer decides one's action from the set-valued signal based on his/her own preference.
%To achieve the reference while allowing the consumers' selfish decision, 

The remainder of this paper is organized as follows. 
Section 2 presents the problem setting of the CEMS design. 
Section 3 is devoted to the controller design and the analysis of the overall CEMS. 
In Section 4, a demonstration of the CEMS is given. 
Section 5 gives the conclusion of this paper.

Notation:  
The symbol ${\bf 1}_n$ represents the all-one vector defined in $\Rn$.
For any signal $x(t)$, the symbol $x(s)$ denotes the Laplace transformation.  
%The symbol $L_2$ represents the set of $L_2$-bounded signals.
For any $L_2$-signal $x(t)$, which is denoted by $x(t) \in L_2$, the symbol $\|x\|_{L_2}$ represents the $L_2$ norm.
For any $L_2$-stable system $G$, the symbol $\| G \|_{L_2}$ represents the $L_2$ gain.
%The symbol $RH_\infty$ represents the set of rational, proper, and stable transfer matrices.
For any rational, proper, and stable transfer matrix $G(s)$, which is denoted by $G(s) \in RH_\infty$, 
the symbol $\| G(s) \|_{H_\infty}$ represents the $H_\infty$ norm.  It holds that $\|G(s)\|_{H_\infty} = \|G\|_{L_2}$.

%=============================================================================
%=============================================================================
\section{Problem Setting}%Section2

\begin{figure}[t]
\begin{center}
\includegraphics[width=\linewidth]{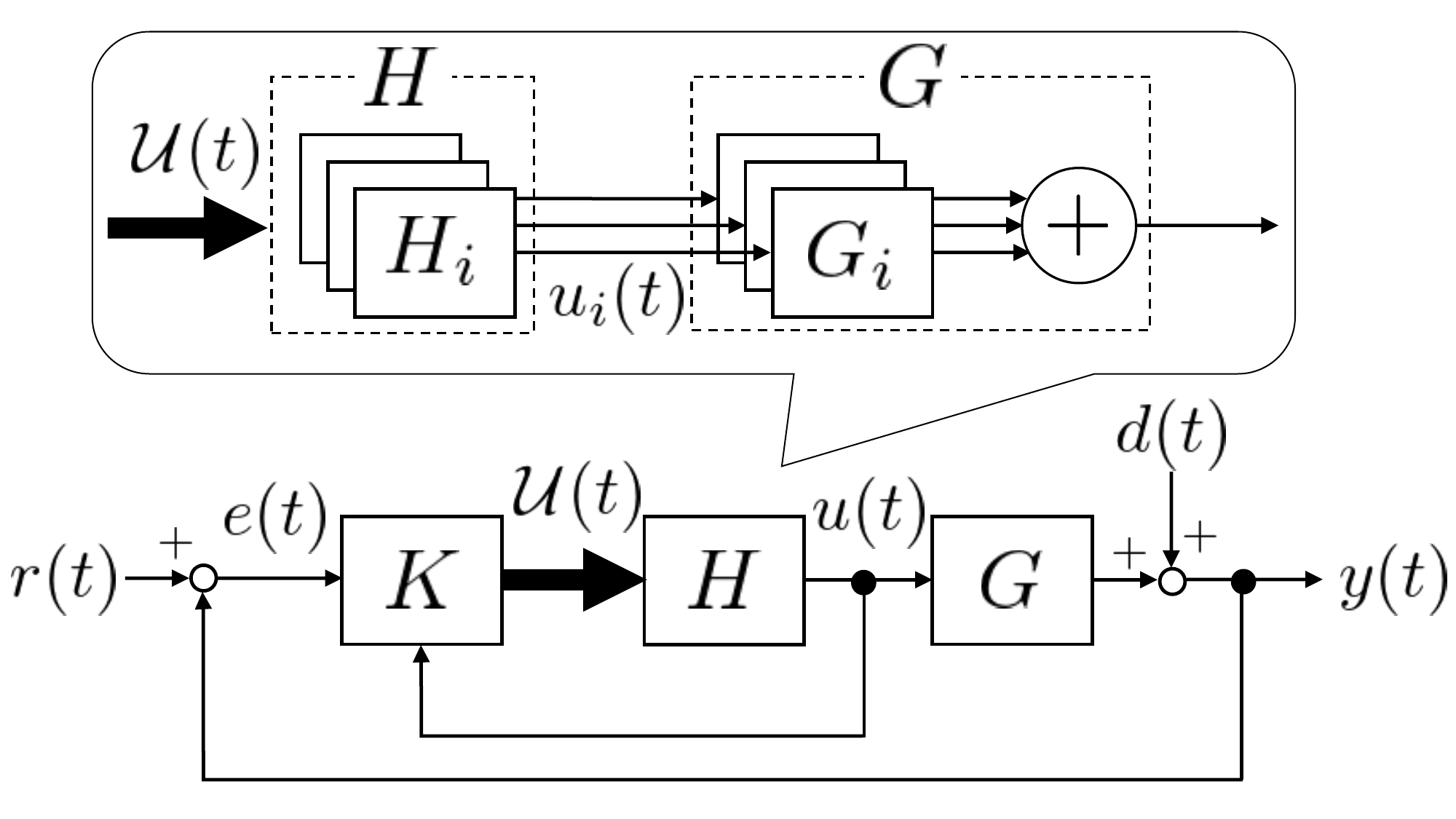}    % The printed column width is 8.4 cm.
\caption{Overall structure of CEMS} 
\label{block1}
\end{center}
\end{figure}

In this section, we consider the CEMS that includes the decision making and actions by consumers.
Fig.~\ref{block1} shows the block diagram of the overall structure of the CEMS.
In the figure, $G$, $H$, and $K$ represent the plant set, consumer set, and controller, respectively.
Note that $G$ and $H$ contain various plant systems $G_i, i \in \{1,2,\ldots, n\}$ and 
various consumers $H_i, i \in \{1,2,\ldots, n\}$, respectively.  Each consumer $H_i$ is responsible for controlling his/her own plant $G_i$.
In this paper, we address the design problem of $K$ for given $G$ and $H$.
The model of $G$, $H$, and $K$ are given as follows.

The plant set $G$ is a dynamical system, and it determines the actual power-saving amount, denoted by $y(t) \in \mathbb{R}$, based on the control action for power-saving, denoted by $u(t) \in \mathbb{R}^{n}$. 
In addition, let $d(t) \in \mathbb{R}$ denote the disturbance to the power-saving amount. 
Then, the model of $G$ is described by the following transfer function representation:
%%%
\begin{align}
G:~y(s)=G(s)u(s)+d(s), \label{eq1}
\end{align}
%%%
where  $G(s)$ is the transfer  matrix given by
%%%
\begin{align}
G(s) = [\,G_1(s)\ G_2(s)\ \cdots\ G_n(s)\, ] \label{eq2}
\end{align}
%%%
and each of $G_i(s)$, $i \in \{1,2,\ldots, n\}$ represents the transfer function of $G_i$.
For simplicity, $y(t)$, $u(t) =[\, u_1(t) \cdots u_n(t)\,]^\top$, and $d(t)$ are called the output, action, and disturbance, respectively.
We consider that each $G_i$ is composed of electrical equipments and that the action $u_i(t)$ is to provide the set point of power-saving amount to $G_i$. 
Furthermore, it is assumed that the output of each $G_i$ tracks the set point at the steady state.  
Then, $G(s)\in RH_\infty$, and the following technical assumption on the steady-state property is imposed on $G$.

\begin{assu}
\label{ass:1}
It holds that $G(0) = {\bf 1}_n^\top$. 
\end{assu}

The consumer set $H$ decides the action $u(t)$ based on the set of allowable control actions $\Ucal(t) \subset \mathbb{R}^{n}$, 
which is requested from the controller $K$ and is modeled by a {\it set-valued} signal.  
The allowable set $\Ucal(t)$ is called the request in the remainder of this paper.
The concept of controlling decision makers by providing the set-valued control signal is called {\it weak} control (see the original work by \cite{inoue2019weak}).  
For any time $t$, the decision made by $H$ is modeled by  the following optimization:
%%%
\begin{align}
H: \left\{\begin{array}{ll}
\underset{u(t)}{\rm{min}} \quad & f(t, u(t)), \\
{\rm subject~to} \quad & u(t) \in \Ucal(t), \\
~ & g_1(t, u(t)) = 0, \\
~ & g_2(t, u(t)) \leq 0. \\
\end{array}\right.
\label{eq24}
\end{align}
%%%
Note here that the functions $f(t, u(t))$, $g_1(t, u(t))$, and $g_2(t, u(t))$ can be time-varying and {\it private}, i.e., they are not open to controller designers or system managers.
The key of the optimization problem is the constraint
\begin{align}
	u(t) \in \Ucal(t), \label{eq31}
\end{align}
%%%
which is the {\it rule} imposed on the consumers.
As long as the consumers follow the request such that \eqref{eq31} holds,
 they can pursue their own benefits by minimizing the cost $f(t, u(t))$. 
The minimization implies reducing their physical/mental burden caused by power-saving.
It is assumed that for any $\Ucal(t)$, the optimization problem \eqref{eq24} is feasible for some $u(t)$.

Let $r(t) \in \mathbb{R}$ and $e(t) \in \mathbb{R}$ denote the reference of the power-saving amount and the tracking error defined by $e(t) :=r(t)-y(t)$, respectively.
Then, the model of $K$ is described by
%%%
\begin{align}
K:~\mathcal{U}(t) = K(e(t), u(t)), \label{eq4}
\end{align}
%%%
where $K(e(t),u(t))$ represents a dynamical system and its details are given in Section 3.  
It is implicitly assumed that the action $u(t)$ made by $H$ is available in the operation of the controller $K$.
This assumption plays a key role in the stability assurance of the overall CEMS, which is stated in Subsection 3.1.

This paper addresses the design problem of the controller $K$ pursuing the following two aims;
1) the accurate reference tracking for the power-saving amount under the presence of disturbance, 
2) with (partially) allowing consumers their selfish actions. 
Let $G_{yr}$ and $G_{yd}$ denote the input-output dynamical systems from $r(t)$ to $y(t)$ and from $d(t)$ to $y(t)$, respectively.
Then, Aim 1 is reduced ultimately to $G_{yr}=1$ and $G_{yd}=0$.
Aim 2 is modeled in \eqref{eq24}, where the consumers aim at minimizing the cost function {\it without taking care} of the  stability or the control performance of the overall CEMS.

Note that pursuing Aim 1 is beneficial to the CEMS manager, while Aim 2 is clearly beneficial to consumers.  
In a point of view of the CEMS manager,
the model of $H$, including $f(t, u(t))$ and $g_j(t, u(t))$, is unavailable for any of the design, implementation, and operation of $K$. 
Then, the design problem of $K$ pursuing Aims 1 and 2 is formulated in the following problem.

\begin{problem}
\label{prob:1}
Given  $G$, find $K$ such that the following two statements hold independently of $H$.
\begin{itemize}
\item The feedback system composed of $G$, $H$, and $K$ is stable. 
\item Given specific $\rho$, it holds that $\| y \|_{L_2} \leq \rho$ { under the presence of $d(t)$}.
\end{itemize}
\end{problem}

This section states the problem of {\it weakly} controlling the consumer set $H$ by providing the request $\Ucal(t)$.
By the weak control, we aim at both of the stability assurance and control performance under the presence of selfish actions.

%=============================================================================
\section{Controller Design}\label{3}%Section3

\begin{figure}[tb]
\centering
\includegraphics[width=\linewidth]{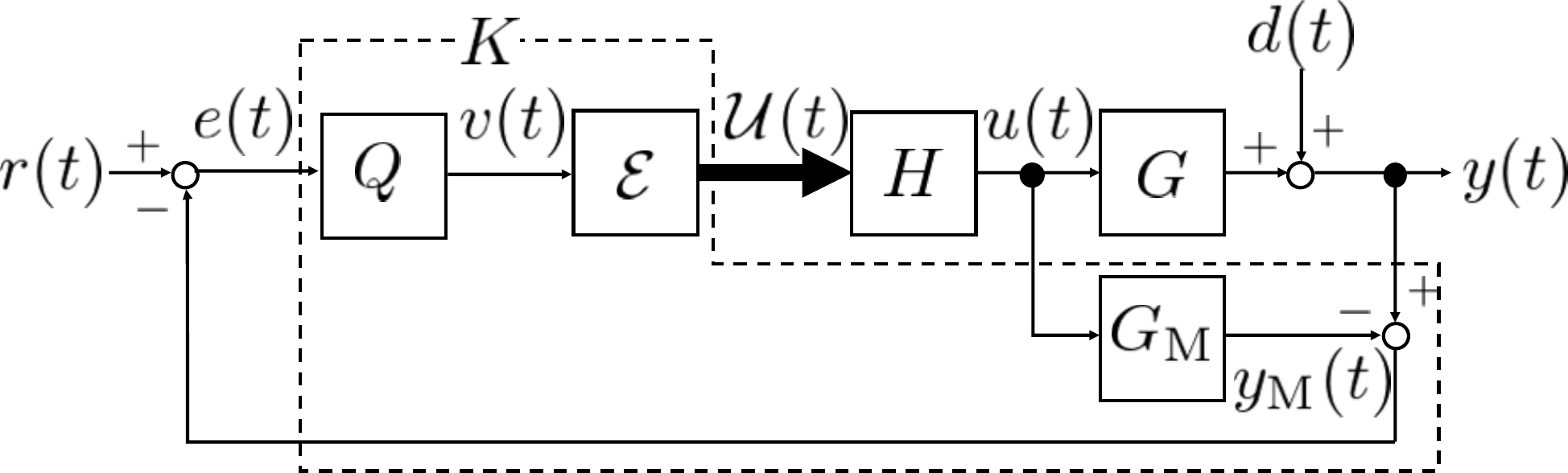}
\caption{Block diagram of controller $K$}
\label{block_wE}
\end{figure}

\subsection{Controller Structure for Stability Assurance} \label{3.1}%==================
We propose the IMC-based structure in the controller $K$, which is illustrated in \rfig{block_wE}, to solve Problem~\ref{prob:1}.
In the figure, the controller $K$ is composed of the  plant model $G_{\rm M}$, filter $Q$, and expander $\Ecal$,
which are described as follows.

The plant model $G_{\rm M}$ is described by 
%%%
\begin{align}
	G_{\rm M}:\ { y_{\rm M}(s)} = G_{\rm M}(s) u(s), 
\end{align}
%%%
where $y_{\rm M}(t) \in \mathbb{R}$ is the model output,  $G_{\rm M}(s)$ is the transfer matrix given by
%%%
\begin{align}
	G_{\rm M}(s) =[\,G_{{\rm M}1}(s)\ G_{{\rm M}2}(s)\ \cdots\ G_{{\rm M}n}(s)\,],
\end{align}
%%%
and each $G_{{\rm M}i}(s)$ represents the model of $G_i$.

The filter $Q$ is called the Youla parameter after the pioneering work by \cite{Youla_76part1}.
The input-output dynamics of $Q$ is described by 
%%%
\begin{align}
	Q:\ v(s) = Q(s) e(s),
\end{align}
%%%
where $v(t) \in \mathbb{R}$ is called the filtered reference and $Q(s)$ is the transfer function satisfying $Q(s) \in RH_\infty$.

\begin{figure}[tb]
\centering
\includegraphics[width=4cm]{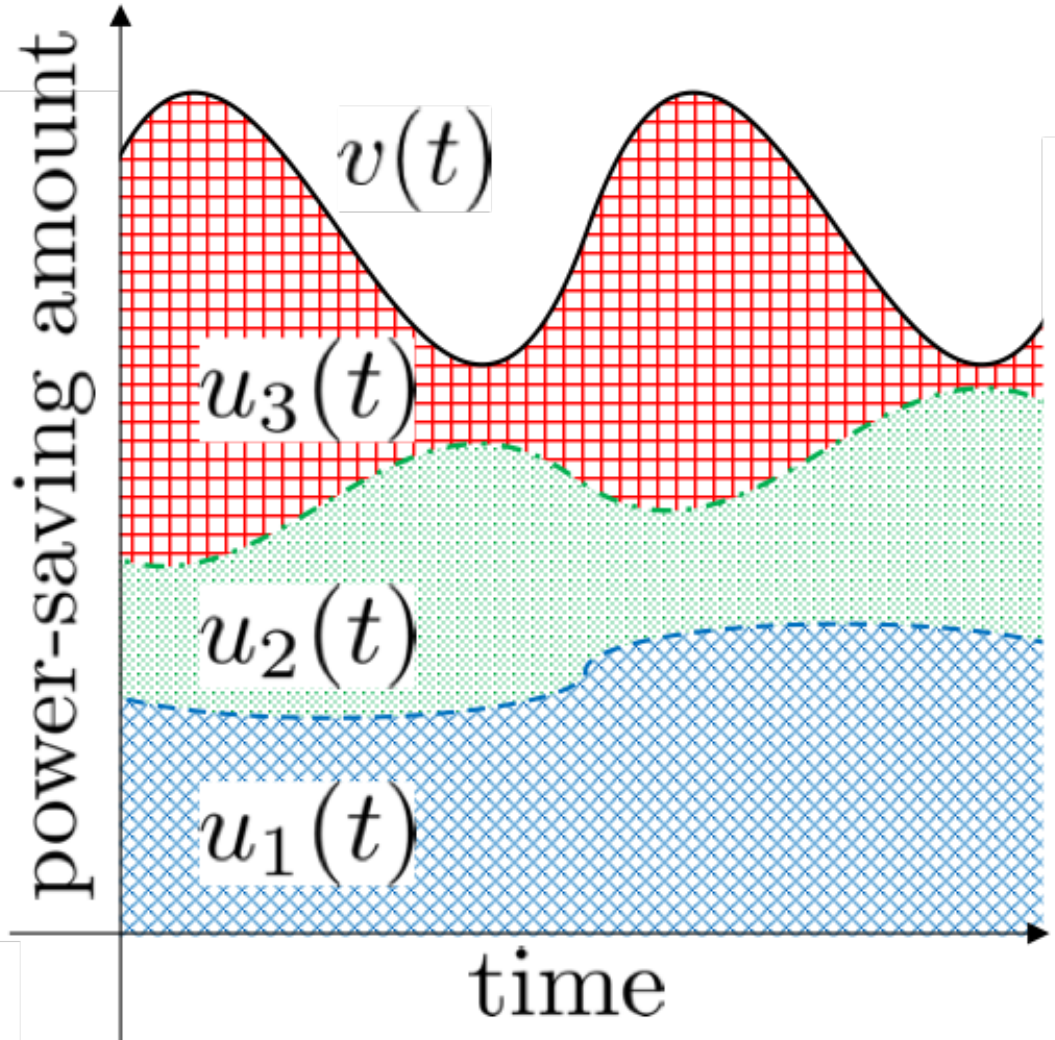}
\caption{Resource allocation in power-saving}
\label{dist}
\end{figure}

%Ecal
The expander $\Ecal$ plays a central role of allowing $H$ selfish actions
by generating the set-valued signal $\Ucal(t)$ based on $v(t)$.
The input-output behavior of $\Ecal$ is described in the time-domain as
%%%
\begin{align}
\Ecal:\ \mathcal{U}(t) = \Ecal(v(t)), \label{eq10}
\end{align}
%%%
where $\Ecal(v)$ is the set-valued function given by
%%%
\begin{align}
\Ecal(v)=\left\{ \left.\left[
\begin{array}{c} u_1 \\\vdots \\  u_n\end{array}
\right]\right|
\sum_{i=1}^{n} {{u}_i} = v,~
{{u}_i} - \frac{1}{n} v \in [ -\frac{\gamma_{{\rm l},i}}{n}v, \frac{\gamma_{{\rm u},i}}{n}v]
\right\}
 \label{eq11}
\end{align}
%%%
and $\gamma_{{\rm l},i}$ and $\gamma_{{\rm u},i}$ are positive constants.
The center of the set-valued signal $\Ucal(t) = \Ecal(v(t))$ is given by $\frac{1}{n}v(t){\bf 1}_n$, and 
its volume, i.e., the degree of freedom, is characterized by the values of $\gamma_{{\rm l},i} $ and $\gamma_{{\rm u},i}$.
Note that the constraint $\sum u_i=v$  means the {\it resource allocation}: 
the resource $v(t)$ is allocated to the consumers for their $u_i(t)$.
The allocation for the case $n = 3$ is illustrated in \rfig{dist}.
The total power-saving amount of the consumers is equal to the filtered reference $v(t)$ generated by the filter $Q$.

The design problem of the controller $K$, which is formulated in Section 2, is reduced to the design of $Q$ and $\Ecal$.
The most fundamental theorem on the stability of the overall system is given. 
%The following theorem states that 
%any stable and proper $Q$ guarantees the stability.
%and With the presence of $\Ecal(v(t))$, following {\it Proposition~1} holds (\cite{inoue2019weak}).
\smallskip

\begin{theo}
\label{theo:1}{\it (Stability):}
Suppose that $Q(s)\in RH_\infty$ and $G_{\rm M}(s)=G(s)$.  
Then, the feedback system composed of $G$, $H$, and $K$ is $L_2$-stable independently of the model (\ref{eq24}).
\end{theo}
\smallskip

\begin{figure}[tb]
\centering
\includegraphics[width=0.85\linewidth]{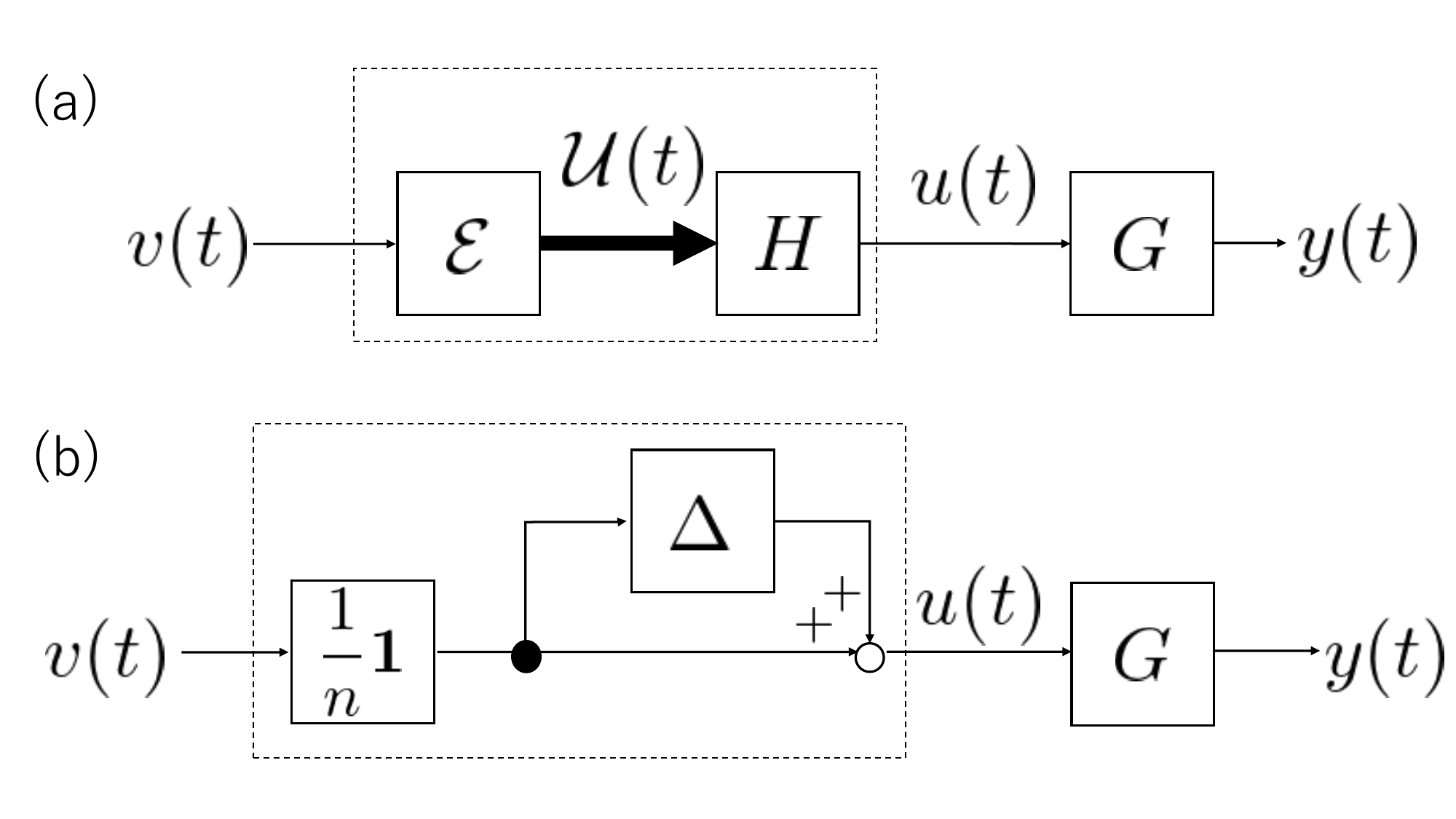}
\caption{Transformation of $\Ecal$ and $H$}
\label{block_ex}
\end{figure}

{\it Outline of Proof of Theorem \ref{theo:1}.}\quad
{ 
%Only the outline of the proof is stated in this paper.
First, recall that $\frac{1}{n}v(t){\bf 1}_n$ is the center of $\Ucal(t)$.
Let $\Delta_i$ be the operator that outputs the error between $u_i(t)$ and $\frac{1}{n}v(t)$ based on the input $\frac{1}{n}v(t)$.
Furthermore, we assume that $\Delta_i$ is in the class of the linear time-invariant (LTI) systems for simplicity of notation.%
\footnote{In the proof of Theorem \ref{theo:1} and Proposition \ref{prop:2}, 
it is additionally assumed that each $\Delta_i$ is LTI.
This {\it technical} assumption is imposed such that the redundant operator representation of subsystems is avoided. 
Details are omitted in this paper, but the assumption is not necessary for the statements of the theorem and proposition.
%The complete proof with no assumption is obtained in a similar manner to the work by \cite{inoue2019weak}.
}  
Then, we have the expression
%%%
\begin{align}
	u_i(s) - \frac{1}{n}v(s) =\Delta_i(s) \left(\frac{1}{n}v(s)\right),
	\label{eq:proofTheo1}
\end{align}
%%%
where $\Delta_i(s)$ is the transfer function of $\Delta_i$.
Further letting
%%%
\begin{align}
	\Delta(s)=\rm{diag}\{\Delta_i(s)\}, \label{eq42}
\end{align}
%%%
it follows that
the input-output system from $v(t)$ to $u(t)$ is written by
%%%
\begin{align}
u(s)&=({I} + \Delta(s)) \left(\frac{1}{n}{\bf 1}_n v(s)\right).
%&=\frac{1}{n}\left[\begin{array}{c}1+\Delta_1(s) \\
%\vdots \\
%1+\Delta_n(s)\end{array}\right] v(s).
\label{eq34}
\end{align}
%%%
The derivation of the expression \eqref{eq34} is illustrated in \rfig{block_ex} (a) and (b).

Next, suppose $G_{\rm M}(s)=G(s)$.
Then, the overall feedback system is described by
%%%
\begin{align}
y(s)=&\frac{1}{n}G(s)(I+\Delta(s)){\bf 1}_n Q(s) r(s)\\
&+\left( 1-\frac{1}{n}G(s)(I+\Delta(s)){\bf 1}_n Q(s)\right)d(s).
\label{eq33}
%&=G_{yr}'(s)r(s)+G_{yd}'(s)d(s). \label{eq41}
\end{align}
%%%
The structure of \eqref{eq33} is the cascaded one, i.e., $G$, $\Delta$, and $Q$ are connected in serial. 
Note that $\Delta$ is $L_2$-stable since $\|\Delta\|_{L_2} (= \|\Delta(s)\|_{H_\infty})$ is 
bounded by $\max\{\gamma_{{\rm l},i}, \gamma_{{\rm u},i}\}$ from the definition of $\Ecal(v)$ in \eqref{eq11}.
Then, noting $G(s)$, $Q(s) \in RH_\infty$, we show that the overall feedback system is $L_2$-stable.  
%Therefore, the overall system is $L_2$-stable.
%In a more general case on $\Delta$, i.e., $\Delta$ is nonlinear or time-varying, this result can be derived in a similar manner.  
This completes the proof.
\hfill \qed
}

\smallskip

\begin{rema}
Theorem \ref{theo:1} claims that the stability of the overall control system is guaranteed even if any action is made by $H$.
It should be emphasized that the stability is also independent of the values of $\gamma_{{\rm l},i}$ and $\gamma_{{\rm u},i}$, 
which determine the volume of the decision space in $H$.
\end{rema}

\subsection{Design of Youla Parameter $Q$} \label{3.2}%=======================================
%\UTF{0092}i\UTF{0097}\UTF{008E}_Ecal\UTF{0093}\UTF{0099}\UTF{0095}\UTF{00AA}\UTF{0094}z
We propose the design strategy of $Q$ to solve Problem~\ref{prob:1} in addition to ensuring the stability of the overall feedback system.
We first briefly review the IMC-based parameter design.  
Then, we show that the design is also applicable to the weak control problem, stated in Problem \ref{prob:1}.

To begin with, we suppose that  $\gamma_{{\rm l},i}=\gamma_{{\rm u},i}=0$ holds for all $i \in \{1,2,\ldots,n\}$ in \eqref{eq11}.
Then,  $\Ucal(t)$ is described by
\begin{align}
\Ucal(t) = \dfrac{1}{n}{\bf 1}_n v(t). \label{eq12}
\end{align}
%%%%
This implies that $v(t)$ is equally distributed to $u_i(t)$, 
i.e., $u_i(t) = \frac{1}{n} v(t)$ holds for all $i \in \{1,2,\ldots,n\}$. 
Let $G_{yr}(s)$ and $G_{yd}(s)$ denote the transfer functions from $r(t)$ to $y(t)$ and from $d(t)$ to $y(t)$, respectively.
Then, noting that $\Delta = 0$ holds in (\ref{eq33}), it follows that
%%%
\begin{align}
G_{yr}(s)&=\frac{1}{n}G(s){\bf 1}_n Q(s), \label{eq:Gyr}\\
G_{yd}(s)&=1-  \frac{1}{n}G(s){\bf 1}_n Q(s). \label{eq:Gyd}
\end{align}
%%%
On the basis of the classical IMC approach  by \cite{morari1989robust}, 
$Q$ is designed by
\begin{align}
Q(s) = \frac{nF(s)}{G_{{\rm M}}(s){\bf 1}_n}, \label{eq18}
\end{align}
%%%
where $F(s)$ is any filter system satisfying $F(0)=1$.
It follows that $G_{yr}(s) = F(s)$ and $G_{yd}(s) = 1- F(s)$ hold.
This implies that the perfect tracking is achieved at the steady-state, i.e., 
for the step reference $r(t)\equiv r_0$, it holds that $y(t) \rightarrow r_0$, $t \rightarrow \infty$.

From now on, the weak control problem is addressed.
In other words, $\gamma_{{\rm l},i}\neq \gamma_{{\rm u},i}$ holds for some $i$ in \eqref{eq11}.
Although the control performance may be deteriorated at the transient state due to the selfish actions made by $H$, 
the stability of the overall system is guaranteed independently of the actions as discussed in Subsection 3.1.
In addition, it should be emphasized that the control performance at the steady state is NOT deteriorated compared with the case of the equal distribution (\ref{eq12}).
This fact is summarized in the following proposition.
\smallskip

%prop2
\begin{prop}
\label{prop:1}{\it (Steady State Performance):}
Suppose that $Q(s) \in RH_\infty$ is designed by (\ref{eq18}),  $G_{\rm M}(s) = G(s)$, and $d(t)\rightarrow d_0, t \rightarrow \infty$ for some $d_0$.
Then, under Assumption \ref{ass:1}, 
 for any step reference $r(t)\equiv r_0$ it holds that $y(t) \rightarrow r_0$, $t \rightarrow \infty$ independently of the model (\ref{eq24}).
\end{prop}
\smallskip

{\it Proof of Proposition \ref{prop:1}.}\quad
At the steady state, we see that 
\begin{align}
y(t) &= G(0) u(t) + d_0,\\
v(t) &=Q(0) (r(t) - d_0)
\end{align}
%%%
hold. 
From Assumption \ref{ass:1} and (\ref{eq11}), it holds that 
\begin{align}
G(0) u(t) = {\bf 1}_n^\top u(t) = v(t).
\end{align}
%%%
In addition, noting that $Q(0) = F(0) = 1$ holds, 
 we show that $y(t)\rightarrow r_0$, $t \rightarrow \infty$ holds.
This completes the proof of the proposition.
\hfill \qed
\smallskip

\begin{rema}
Proposition \ref{prop:1} claims that the {\it perfect tracking} to the step reference is guaranteed independently of the actions made by $H$.
In other words, this steady state performance is independent of the values of $\gamma_{{\rm l},i}$ and $\gamma_{{\rm u},i}$, 
while the transient performance depends.
In the next subsection, we address their design  such that the transient error from the {\it nominal} behavior is bounded.
\end{rema}

\subsection{Design of Expander $\Ecal$} \label{3.3}%====================================
In this subsection, the design problem of the expander $\Ecal$ is addressed.
For simplicity of discussion, consider that $r(t) \equiv 0$, i.e., only the disturbance suppression in $y(t)$ is addressed.
In addition, we let 
%%%
\begin{align}
\gamma_i := \gamma_{{\rm u},i} = \gamma_{{\rm l},i},\  i \in \{1,2\ldots,n\}
\end{align}
in \eqref{eq11}.
Then, the design problem of $\Ecal$ is reduced to that of $\gamma_i$, $i \in \{1,2\ldots,n\}$.
From the view of the CEMS manager, the aim of  the design is 
to bound the worst case behavior in $y(t)$ caused by the actions of $H$ and the disturbance $d(t)$.

As a preliminary, we estimate the {\it nominal} behavior, where $\Ucal(t)$ is given by \eqref{eq12} and any selfish action is not allowed for $H$.
Recall the transfer function $G_{yd}(s)$ of \eqref{eq:Gyd}.
Supposing that $Q$ is designed by \eqref{eq18}, we have $G_{yd}(s) = 1- F(s)$.
This results in $y(s) = d(s)- F(s)d(s)$. 
We let $d_{\rm f}(t)$ represent the filtered disturbance and be described in the Laplace domain by 
%%%%
\begin{align}
	d_{\rm f}(s) := F(s) d(s).
\end{align}
%%%
Then, the zero initial state response of $y(t)$ satisfies 
%%%%
\begin{align}
	\|y\|_{L_2} = \|d - d_{\rm f}\|_{L_2},\label{eq:nominal}
\end{align}
%%%
which is the nominal performance for the disturbance suppression.

On the basis of the estimation \eqref{eq:nominal},
we give the estimate of the {\it general} behavior generated in the weak control problem, 
where $\Ucal(t)$ is given by \eqref{eq11} and the selfish action is partially allowed for $H$.
%For simplicity, a technical assumption is imposed such that $\Delta_i$ is an LTI system.  
\smallskip

%prop3
\begin{prop}
\label{prop:2}{\it (Disturbance Suppression):}
Suppose that $Q(s) \in RH_\infty$ is designed by (\ref{eq18}), $G_{\rm M}(s) = G(s)$, $r(t) \equiv 0$, and $d(t) \in L_2$.
Then, letting $\gamma_i$, $i \in \{1,2,\ldots,n\}$ satisfy
%%%%
\begin{align}
\sum_{i = 1}^n \gamma_i \|G_i(s)\|_{H_\infty} \leq 
\frac{\varepsilon n}{ \|Q(s)\|_{H_\infty} \|d\|_{L_2}} \label{eq43}
\end{align}
%%%
for some positive constant $\varepsilon$, it holds that 
%%%
\begin{align}
	\|y\|_{L_2} \leq \|d - d_{\rm f}\|_{L_2} + \varepsilon.
\label{eq44}
\end{align}
\end{prop}
\smallskip

%\UTF{008F}\UTF{00D8}\UTF{0096}\UTF{0178}%
{\it Proof of Proposition \ref{prop:2}.}\quad
Recall the operator $\Delta_i$ from the proof of Theorem \ref{theo:1}.
Here, we assume again  that each $\Delta_i$ is LTI, 
which is not necessary for the proof, but avoids newly introducing redundant notation.
Then, from the definition of $\Ecal$ of \eqref{eq11}, we see that
%%%
\begin{align}
	\|\Delta_i\|_{L_2} = \|\Delta_i(s)\|_{H_\infty}  \leq \gamma_i,\ i \in \{1,2\ldots,n\}
\end{align}
holds. 

Recalling \eqref{eq33}, which is the expression of $y(t)$ in the Laplace domain,  we have 
%%%
\begin{align}
	y(s) = d(s) - G(s) \frac{1}{n} {\bf 1}_n Q(s) d(s) - G(s) \Delta(s) \frac{1}{n} {\bf 1}_n Q(s) d(s).
\end{align}
%%%
Noting that $Q(s)$ is designed by \eqref{eq18}, this equation is reduced to
%%%
\begin{align}
	y(s) = d(s) - d_{\rm f}(s) - G(s) \Delta(s) \frac{1}{n} {\bf 1}_n Q(s) d(s).
\end{align}
%%%
It follows  that 
%%%
\begin{align}
	\| y\|_{L_2} &\leq  \|d - d_{\rm f}\|_{L_2}\\ 
				&\quad + \frac{1}{n} \left(\sum_{i=1}^n\| G_i(s) \Delta_i(s)\|_{H_\infty}\right)  \|Q(s)\|_{H_\infty} \|d\|_{L_2}.
\end{align}
%%%
We see from \eqref{eq43} that \eqref{eq44} holds.  This completes the proof of the proposition. 
\hfill \qed

\smallskip

\begin{rema}
In Proposition \ref{prop:2}, the performance bound of $y(t)$ is studied with respect to the designable parameters $\gamma_i$, $i \in \{1,2,\ldots,n\}$.
%Suppose the expander $\Ecal$ is designed such that \eqref{eq43} holds, 
The proposition states that 
the performance deterioration caused by any selfish actions of $H$ is bounded by $\varepsilon$. 
On the basis of the criterion \eqref{eq43}, the CEMS manager can design  $\Ecal$, i.e., equivalently  $\gamma_i$, $i \in \{1,2,\ldots,n\}$.
\end{rema}

%=============================================================================
\section{Numerical Demonstration}\label{4}%Section4
In this section, we demonstrate the design of the controller $K$ to construct the consumer-preferred CEMS.
First, we give the experimental conditions including the description of the plant set $G$ and consumer set $H$. 
Next, for given $G$ and $H$, the controller $K$ is designed, and its effectiveness is shown.
%First, we define the plant system and other experimental conditions.
%Next, the controller $K$ is designed and applied to the model.

\subsection{Experimental Condition} \label{4.1}%=========================================
\begin{figure}[tb]
\centering
\includegraphics[width=\linewidth]{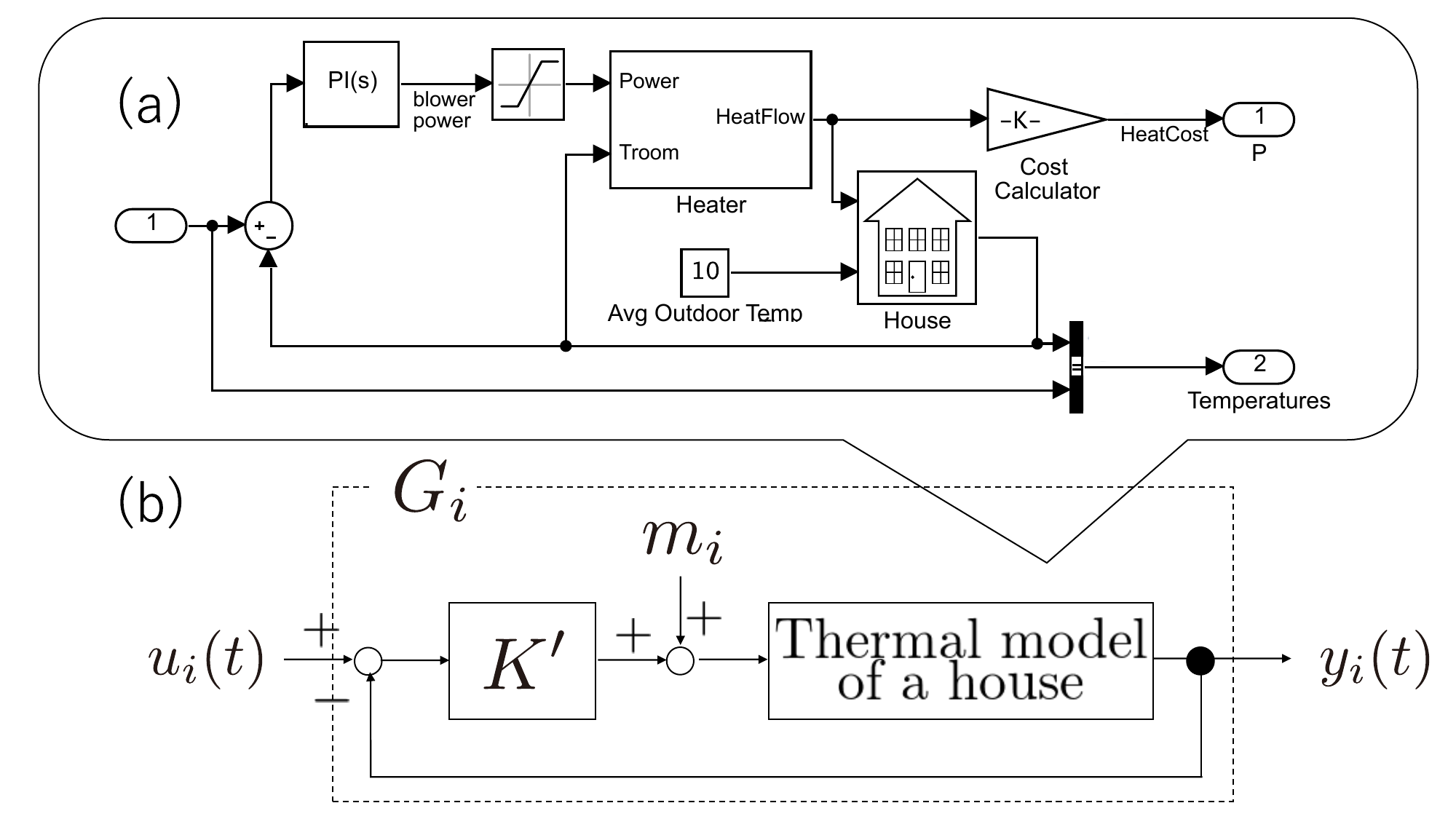}
\caption{Structure of plant $G_i$}
\label{block_house}
\end{figure}

Each plant $G_i, i \in \{1,\ldots, 5\} $ represents an  air conditioning system installed in a house.
A part of the system is described by the thermal model in \cite{thermalmodel}, MATLAB\textregistered~and  is illustrated in { \rfig{block_house}(a)}.
In addition, a local controller $K^\prime$, which is common for all plant systems, is equipped with the thermal model. 
Then, a feedback  system is constructed as illustrated in \rfig{block_house}(b).
The feedback system $G_i$ receives the signal of the power-saving action $u_i(t)$ and generates the power-saving amount {$y_i(t)$}.
%each house (or plant?) G_i consumes power m_i at a steady state (to achieve comfortable temperature).we define m_i as the initial power consumption.
The initial state of each house is different from each other, and 
we let $m_i$ be the power consumption of each house at the initial state.

The consumer set $H$ decides the action $u(t)$ based on the request $\Ucal(t)$.
Let 
%%%
\begin{align}
	c_i(u_i):=u_i^2+{6}iu_i
\end{align}
%%%
be the private cost function of each consumer $H_i$.
Then, letting $m:=[\,m_1\cdots m_5\,]^\top$, the model of $H$ is described as follows.
%%%%
\begin{align}
H: \left\{\begin{array}{ll}
\underset{u}{\rm{min}} \quad
&f(t,u(t)) := \sum_{i=1}^{5} c_i(u_i(t)),\\
{\rm subject~to} \quad & u(t) \in \Ucal(t), \\
~ & g_2(t,u(t)) := 
		\left[\begin{array}{c}
			{u(t)- 0.2m } \\
			{-u(t)}
	\end{array}\right]
	\leq 0. \\
\end{array}\right.\label{eq23}
\end{align}
%%%
%which is given by replacing $f(u)=\sum_{i=1}^{5} c_i(u_i):=u_i^2+10iu_i$, and ignoring $g_1(u)$ and $g_2(u)$ in \eqref{eq24}.
The increase of the value of $c_i(u_i)$ expresses the high burden imposed on each consumer.
Hence, a smaller value of $f(t,u(t))$ is preferred for the consumers.  
The constraint $g_2(t,u(t)) \leq 0$ means that the action of the power-saving by 
$H_i$, denoted by $u_i(t)$,  is limited in 20~\% of the initial power consumption $m_i$ or less.

We consider that some consumers who do not participate in CEMS affect the total power consumption.
Their effects to the total power-saving amount is modeled by the disturbance $d(t)$.
In this demonstration, the disturbance is given by the {\it filtered} normal random number with the average of 0 and the variance of 10. 
The filter $F_d(s)$ is given by $F_d(s)={1}/({10s+1})$.

In this demonstration, we consider two cases;
%\UTF{0093}\UTF{0099}\UTF{0095}\UTF{00AA}\UTF{0094}z
A) the decision making and selfish actions are NOT allowed for $H$, and the power-saving request is equally distributed, i.e., $\Ucal(t)$ is given by \eqref{eq12}, 
and 
%\UTF{008D}\UTF{00C5}\UTF{0093}K\UTF{0089}\UTF{00BB}%
B) the decision making and selfish actions are partially allowed for $H$, and the request from the controller $K$ includes some degree of freedom,
i.e., $\Ucal(t)$ is given by \eqref{eq11}.

\subsection{Design of Controller $K$} \label{4.2}%========================================
%G_M
The controller $K$ is designed based on the IMC structure  as illustrated in \rfig{block_wE}. 
Then, $K$ is composed of $G_{\rm M}$, $Q$, and  $\Ecal$. 

Initially, we obtained the plant model $G_{\rm M}$ by a system identification experiment where
the step response experiment was performed for each $G_{{\rm M}i}$ independently.
We found that every output followed the step input immediately. 
Hence, we obtained the simplistic plant model $G_{{\rm M}i}(s)=1, i \in \{1,\ldots, 5\}$ in this demonstration.

In addition, we designed the low-pass filter $F(s)$ as
%%%
\begin{align}
F(s)&=\frac{1}{1.5s+1}. \label{eq20}
\end{align}
%%%
By utilizing $G_{{\rm M}_i}(s) = 1$ and this $F(s)$, the Youla parameter $Q(s)$ was designed by \eqref{eq18}.

%\Ecal
The expander $\Ecal$ was designed as \eqref{eq12} in Case A), while it was designed as 
%%%
\begin{align}
\Ecal(v)=\left\{ \left.\left[
\begin{array}{c}{u_1} \\\vdots \\{u_n}\end{array}
\right]\right|
\sum_{i=1}^{n} {u_i} = v
%{u_i} \in [0, 0.2m_i]
\right\},\label{eq22}
\end{align}
%%%
in Case B).  %Note here that in\eqref{eq22}, the performance deterioration at the transient state was ignored, i.e., $\gamma_{{\rm l},i}$, $\gamma_{{\rm u},i} \rightarrow \infty$ holds. 
%which realizes the distribution of the allowable set of action with some degree of free aforementioned in case B).
%which indicates the case B).

\subsection{Result} \label{4.3}%=====================================================

%FF/FB比è¼?%FB
\begin{figure}[t]
\centering
\includegraphics[width=0.8\linewidth]{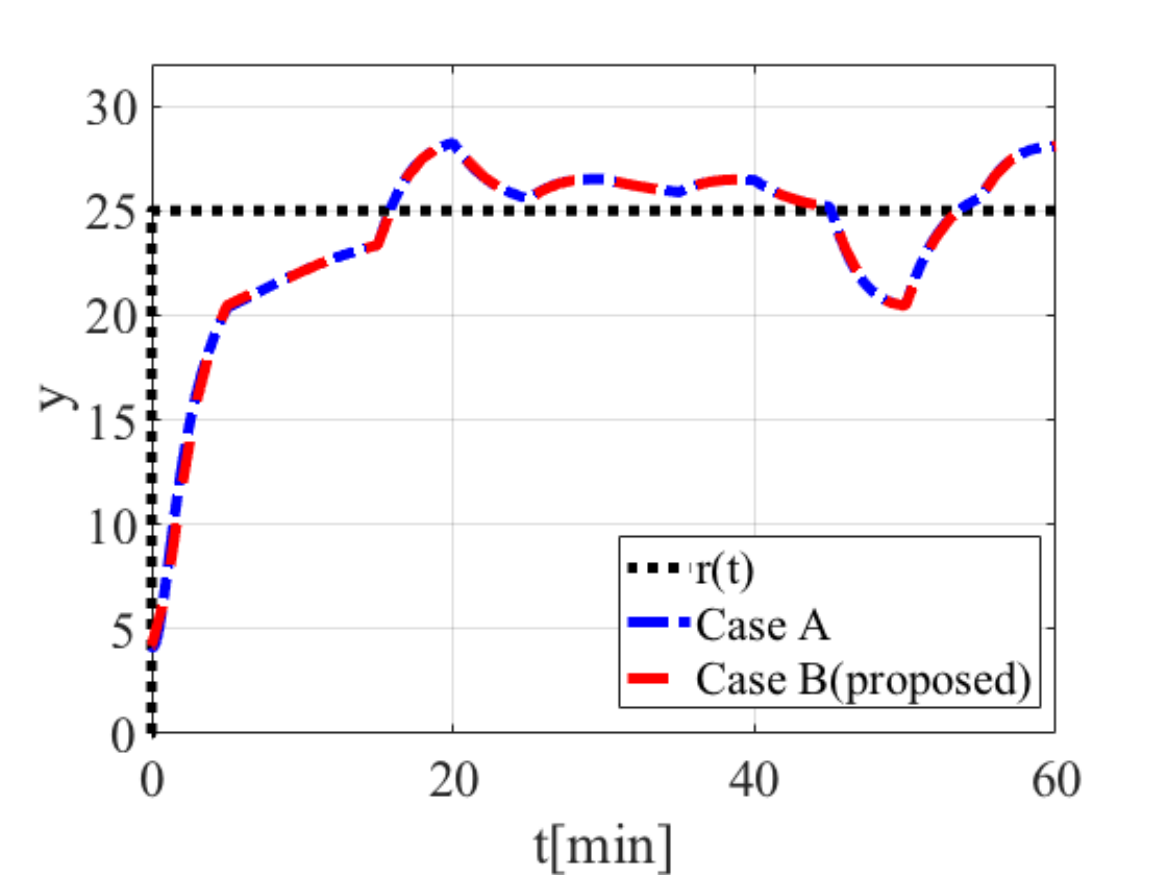}%, height=5cm
\caption{Total power-saving amount achieved in Cases A) and B)}
\label{result_FB_y}
\end{figure}

The results of the power-saving experiments by applying the two controllers A) and B) are illustrated in \rfig{result_FB_y}.
The black dotted, red chained, and blue broken lines represent the reference and the resulting total power-saving amount in Cases A) and B), respectively.
As illustrated in the figure, there is no significant difference in the performance for the reference tracking and disturbance suppression.
To see this fact,  the tracking error $y(t)- r(t)$ is evaluated for Cases A) and B).  Letting $k$ denote the discrete time, it holds that 
%%%
\begin{align}
\sqrt{\sum_{k=1}^{60} (y(k)-r(k))^2} = 36.723
\end{align}
%%%
for the both cases.  This concludes that the selfish decision by the consumers does not deteriorate the overall tracking performance in the CEMS.

%ç¯?電量å?配比è¼?%等å?é…?

\begin{figure}[t]
\centering
\includegraphics[width=0.8\linewidth]{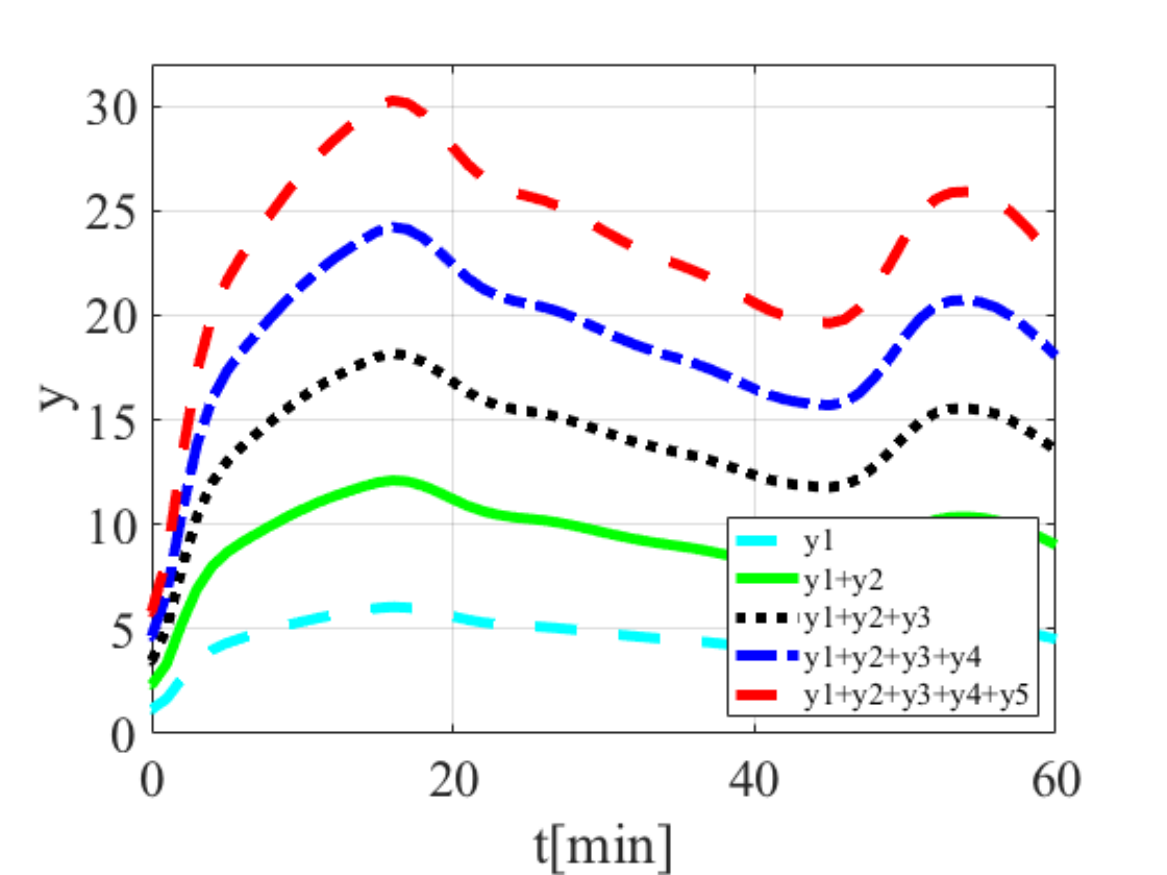}%, height=5cm
\caption{Power-saving amount of each consumer in Case A)
}
\label{result_FB_eq_each}
\end{figure}

\begin{figure}[t]
\centering
\includegraphics[width=0.8\linewidth]{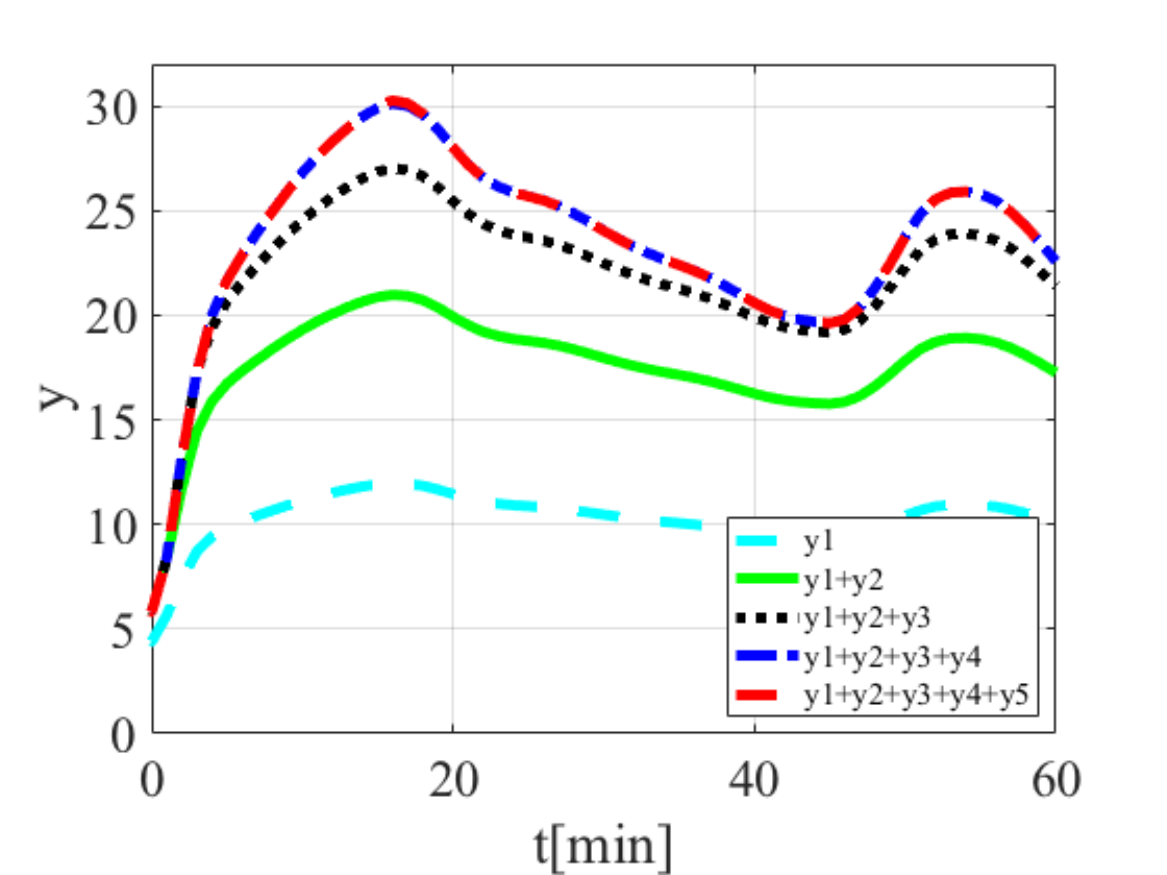}%, height=5cm
\caption{Power-saving amount of each consumer in Case B)}
\label{result_FB_opt_each}
\end{figure}

The actual power-saving amount of each plant $G_i$ in Cases A) and B) are illustrated in 
Figs.~\ref{result_FB_eq_each} and \ref{result_FB_opt_each}, respectively.
Each line shows the actual power-saving amount by the consumers, denoted by $y_1(t)$, $y_1(t) + y_2(t)$, $\ldots$, $\sum_i^5 y_i(t)$.
In other words, the lowest, light blue chained line shows the power-saving amount $y_1(t)$.
The second lowest, green line shows $y_1(t) + y_2(t)$. 
The third lowest, black dotted line shows $y_1(t) + y_2(t) + y_3(t)$, and so on.
In \rfig{result_FB_eq_each}, we see that the power-saving amount is equally distributed to the consumers.
On the other hand, in \rfig{result_FB_opt_each}, the amount is different from each other.
The consumer $H_1$, who has the lowest cost for the power-saving, most contributes to the power-saving, while
$H_5$, who has the higher cost, almost does not contribute.

%æœ?適åŒ?

%コスト比è¼?
\begin{figure}[t]
\centering
\includegraphics[width=0.8\linewidth]{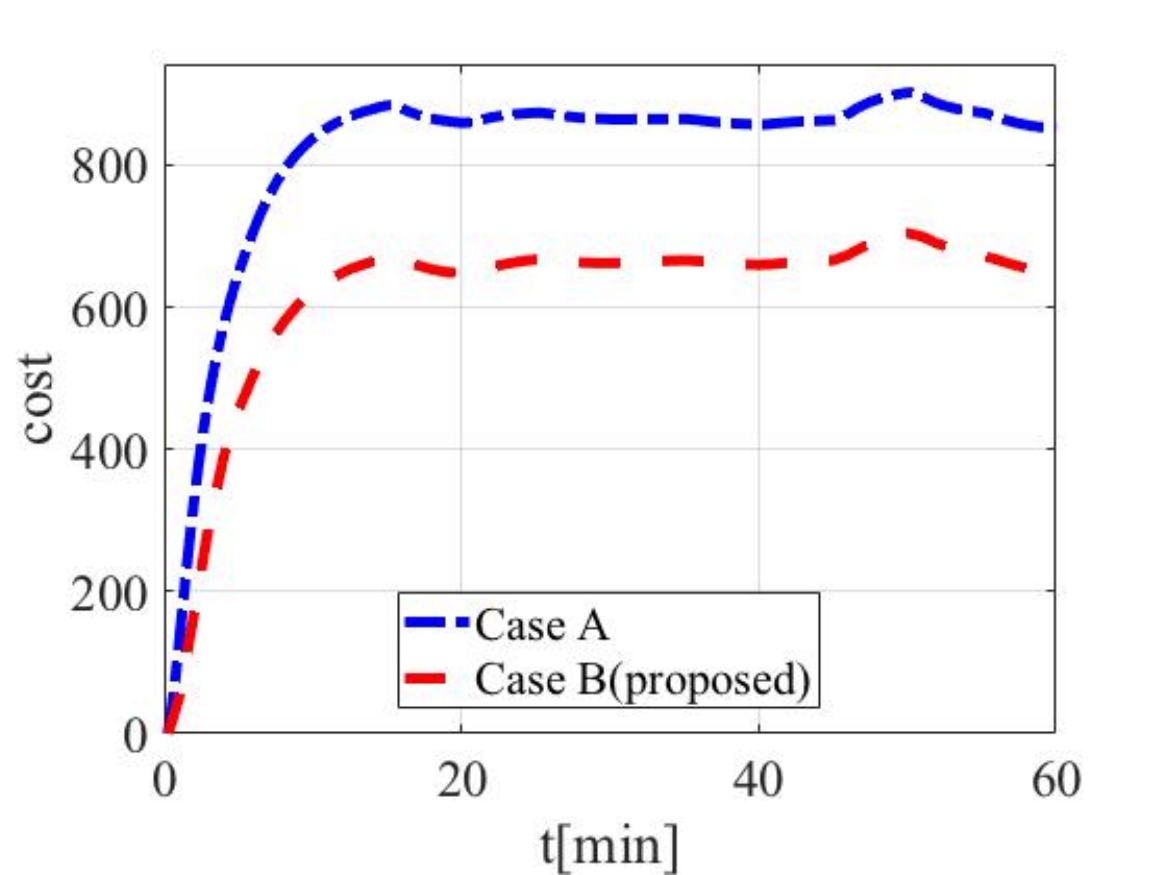}%, height=5cm
\caption{Cost of CEMS achieved in Cases A) and B)}
\label{result_cost}
\end{figure}
%\UTF{008C}\UTF{008B}\UTF{0089}\UTF{00CA}cost
The time trajectory of the total cost $\sum_{i = 1}^{5}c_i({y_i(t)})$ is illustrated in \rfig{result_cost}.
The red chained and blue broken lines represent the trajectory in Cases A) and B), respectively.
This figure shows the total cost is reduced in Case B) compared with Case A).   
We see that the the proposed {\it weak} controller, given in Section \ref{3}, is beneficial for consumers.

%\UTF{0082}\UTF{00DC}\UTF{0082}\UTF{00C6}\UTF{0082}\UTF{00DF}%
This experiment shows that the proposed controller contributes to reducing the private cost of consumers, 
while keeping the control performance within a specified allowable range.

%=============================================================================
\section{Conclusion}%Section5

We proposed the consumer-preferred CEMS, 
in which the system manager partially allowed consumers their selfish actions of power-saving  while achieving desired power-saving amount. 
The design problem of the CEMS controller was formulated and addressed.  Then, the stability, the tracking performance at the steady state, and the disturbance suppression performance at the transient state were studied and stated in Theorem \ref{theo:1} and Propositions \ref{prop:1} and \ref{prop:2}, respectively.  
Finally, a numerical demonstration was performed.  It was shown that the proposed controller is beneficial to consumers in addition to achieving the accurate system management.
 
%The decision maker is designed which decides the power-saving action from the allowable set of action.
%As a result of numerical demonstration,  the cost which means the burden to the whole consumers is lowered and the reference of the power-saving amount is achieved.

%=============================================================================
%\begin{ack}
%Place acknowledgments here.
%\end{ack}

%\bibliographystyle{ifacconf}
\bibliography{IFAC_shiba_10_MIrev}

\end{document}